\UseRawInputEncoding
\documentclass[11pt,nofootinbib,preprint]{revtex4}
\usepackage{mathrsfs}
\usepackage{graphicx}
\usepackage{amsmath}
\usepackage{amsfonts}
\usepackage{amssymb}
\usepackage{color}
\usepackage{hyperref}

\usepackage{epsfig}
\usepackage{CJK}
\usepackage{graphicx}
\usepackage{epsfig}
\usepackage{eepic}
\usepackage{bbm}
\usepackage{dcolumn}
\usepackage{bm}
\usepackage{ulem}
\usepackage{slashbox}
\usepackage{multirow}
\usepackage{slashed}

\newcommand{\omits}[1]{}

\def\bc{\begin{center}}

\def\ec{\end{center}}
\def\be{\begin{eqnarray}}
\def\ee{\end{eqnarray}}

\definecolor{dyellow}{rgb}{1.,0.8,.0}
\definecolor{myblue}{rgb}{.1,.1,.7}
\definecolor{dcyan}{rgb}{.0,.6,.6}
\definecolor{cyan}{rgb}{0.4,1.0,1.0}
\definecolor{dmagenta}{rgb}{0.6,0.0,0.6}
\definecolor{brown}{rgb}{0.6,0.2,0.}
\definecolor{darkblue}{rgb}{.0,.0,0.5}
\definecolor{darkred}{rgb}{0.75,0.0,0.0}
\definecolor{orange}{rgb}{1.,.6,.0}
\definecolor{dorange}{rgb}{0.8,.4,.0}
\definecolor{green}{rgb}{0.0,1.0,0.0}
\definecolor{darkgreen}{rgb}{0.0,0.6,0.0}
\definecolor{purple}{rgb}{.4,.0,.4}
\definecolor{lightgrey}{rgb}{0.7, 0.7, 0.7}
\definecolor{grey}{rgb}{0.4, 0.4, 0.4}

\newcommand{\nc}{\newcommand}
\nc{\rnc}{\renewcommand} \nc{\ket}[1]{\left | \, #1 \right \rangle}
\nc{\bra}[1]{\left \langle #1 \, \right |}
\nc{\ua}{\uparrow} \nc{\da}{\downarrow}

\nc{\braket}[2]{\langle\, #1\,|\,#2\,\rangle}
\nc{\half}{\frac{1}{2}}

\nc{\prj}{\mathcal{P}} \nc{\hilb}{\mathcal{H}}
\nc{\pth}{\mathcal{C}} \nc{\inprod}[2]{\braket{#1}{#2}}
\nc{\upket}{\ket{\uparrow}} \nc{\downket}{\ket{\downarrow}}
\nc{\upbra}{\bra{\uparrow}} \nc{\downbra}{\bra{\downarrow}}

\begin{document}


\title{Distilled density matrices of holographic PEE from thread-state correspondence }

\author{Yi-Yu Lin$^{1,2}$} \email{yiyu@bimsa.cn}

\affiliation{${}^1$Beijing Institute of Mathematical Sciences and Applications (BIMSA),
	Beijing, 101408, China}
\affiliation{${}^2$Yau Mathematical Sciences Center (YMSC), Tsinghua University,
	Beijing, 100084, China}

\begin{abstract}

Within the framework of holographic duality, CMI (conditional mutual information) is often understood as a correlation between ``region pairs" and is closely related to the concept of partial entanglement entropy (PEE). The main theme of this paper is to try to understand the rigorous physical meaning of such a region-pair correlation. This relies on the idea of holographic bit threads and the recently developed thread-state correspondence. In a sense, this effort also prompted us to give a definition of PEE based on the density matrices of the holographic distilled states. Specifically, drawing from experience with the locking multiflow configuration, we first provide a bipartite entanglement explanation for the PEE=CMI scheme, but it leads to difficulties in characterizing the entanglement entropy of disconnected regions. We then introduce multipartite entanglement through the generalized $n$-thread/perfect tensor state correspondence to solve this problem and explain the coincidence between CMI and tripartite information in the holographic quantum systems.

\end{abstract}


\maketitle
\tableofcontents

\section{Introduction}

Entanglement entropy, which characterizes the degree of entanglement in quantum systems, has become an important concept for studying the interesting and profound relationship between quantum entanglement and spacetime geometry. At least in the framework of holographic duality~\cite{maldacena1999large,Gubser:1998bc,Witten:1998qj}, the Ryu-Takayanagi formula states that the holographic entanglement entropy characterizes the information of the area of a minimal surface in the higher-dimensional spacetime~\cite{Ryu:2006bv,Ryu:2006ef,Hubeny:2007xt}. However, generally speaking, given a pure state system and some subsystem $A$, the entanglement entropy $S(A)$ is a highly non-local quantity that characterizes the quantum entanglement between the subregion $A$ and its complement $A_c$. Faced with this property, a natural (maybe naive but prove useful) method is to use the idea of ``whole equals the sum of its parts", and regard the entanglement entropy of $A$ as the sum of contributions from the various parts $A_i$ that make up $A$, subject to the constraints that $\cup {A_i} = A$ and ${A_i} \cap {A_j} = \emptyset $. Indeed, this is the basic idea of the so-called partial entanglement entropy (PEE) proposed in~\cite{vidal2014}. Formally, one can define $f_A(x)$ (named entanglement contour) as a density function of $S(A)$ with respect to the spatial coordinate $x$ of $A$:
\begin{equation}\label{loc}
	S(A) = \int_A {f_A(x)dx}.
\end{equation}
Then the partial entanglement entropy is defined as
\begin{equation}
	s_A(A_i) \equiv \int_{A_i} {f_A(x)dx},
\end{equation}
which represents the contribution of a component $A_i$ in $A$ to the entanglement entropy of $A$. How can we write down the explicit expression of PEE? One way is to constrain it by its physical meaning, requiring it to satisfy a series of reasonable conditions such as additivity, invariance under local unitary transformations, positivity, and permutation symmetry, etc.~\cite{vidal2014,wen2020formulas}. On the other hand, a scheme that can be called PEE=CMI scheme~\cite{Kudler-Flam:2019oru,Wen:2018whg,wen2020entanglement,Lin:2021hqs}\footnote{It is also called the PEE proposal in the early context.}, especially in the framework of holographic principle, has attracted some research attention (see e.g.\cite{wen2020formulas,wen2019towards,wen2020entanglement,han2019entanglement,han2021first,kudler2020negativity,Wen:2018whg} and its interesting application in the study of black hole island problem~\cite{Lin:2022aqf,Rolph:2021nan,ageev2022shaping}). CMI (conditional mutual information) is a quantity in (quantum) information theory used to measure the information correlation involving three objects $X$, $Y$, $Z$. It is defined as
\begin{equation}
	I(X,Y|Z) = S(XZ) + S(YZ) - S(XYZ) - S(Z)
\end{equation}
In particular, when $Z$ is taken to be empty, it reduces to the mutual information $I(X,Y)$. The PEE=CMI scheme suggests that PEE can be given by the expression of CMI. However, the fundamental definition of PEE in terms of the reduced density matrix has not yet been established. This paper is an attempt to explore this direction. The first key point is to realize that, under the framework of PEE=CMI, for analyzing entanglement structure, it is more natural to take pairs of points, rather than the local points (as in~(\ref{loc})) as the basic objects. In other words, the kinematic space perspective~\cite{Czech:2015kbp,Czech:2015qta} is more appropriate than the original space perspective for analyzing the entanglement structure in this context. However, it should be emphasized that this does not mean that PEE or CMI should be understood as a kind of bipartite entanglement. In fact, this is what we want to make clear in this paper.

The insights rely on the recently proposed thread-state perspective, which was proposed independently in~\cite{Lin:2022flo,Lin:2022agc} and~\cite{Harper:2022sky} for different motivations and applications, originating from the study of bit threads~\cite{Freedman:2016zud,Cui:2018dyq,Headrick:2017ucz}~\footnote{For more research on bit threads see e.g.~\cite{harper2018bit,harper2019bit,headrick2022covariant,headrick2022crossing,hubeny2018bulk,lin2021bit,Lin:2022aqf,Lin:2022agc,Lin:2022flo,Lin:2021hqs,agon2019geometric,agon2021bit,agon2021bit2,agon2022quantum,rolph2021quantum,bao2019towards,chen2020quantum,Du:2019emy,Du:2019vwh,Shaghoulian:2022fop,Susskind:2021esx,Pedraza:2021fgp,pedraza2021lorentzian,Harper:2022sky,Harper:2021uuq,Freedman:2016zud,Cui:2018dyq,Headrick:2017ucz,bao2020bit,Kudler-Flam:2019oru}.}. Bit thread is a language that can equivalently describe the RT formula and arose from an analogy with the flow optimization theory in network theory. Within the framework of the holographic principle, let us consider a time slice and liken the bulk manifold to a network, and the boundary to the terminals of the network. We can then consider the flows between two boundary regions, as in a traffic system with flows between cities. Simply put, bit threads are defined as a kind of special unoriented bulk curves subject to a density constraint and, in simple cases, can be described by the field lines of a weak magnetic field. In this language, the entanglement entropies are related to the fluxes in the so-called locking bit thread configurations, due to the max-flow/min-cut theorem in mathematics~\cite{Freedman:2016zud,Cui:2018dyq,Headrick:2017ucz}. In the original PEE=CMI proposal, the entanglement contour has been attempted to be understood as the bit thread density~\cite{Kudler-Flam:2019oru}. In~\cite{Lin:2021hqs}, the concept of multiflow in the bit thread description was further used to more clearly show that partial entanglement entropies can be identified with the component flow fluxes in the locking bit thread configurations, which precisely appear as CMIs. In a sense, this can be viewed as a derivation of the PEE=CMI scheme from a thread perspective. More recently, the physical meaning of bit threads or ``threads" in a more general graph-theoretical sense has been more explicitly explored. Inspired by the entanglement structure between a set of extremal surfaces in the surface/state correspondence~\cite{Miyaji:2015fia,Miyaji:2015yva},~\cite{Lin:2022flo} proposed that each thread in a locking bit thread configuration corresponds to a quantum superposition of two orthogonal states (``red'' state and ``blue'' state). On the other hand, inspired by the holographic entropy cone model~\cite{Bao:2015bfa,Hubeny:2018ijt,Hubeny:2018trv,HernandezCuenca:2019wgh},~\cite{Harper:2022sky} proposed that the so-called $n$-hyperthreads~\cite{Harper:2021uuq} can correspond to perfect tensor states. In fact, for $n=2$, the two proposals are essentially the same. It is natural to ask whether the thread-state correspondence will bring new insights into the holographic PEE scheme. We will answer this question in this paper.

We will show that the thread-state correspondence allows us to provide an explicit definition for the holographic PEE based on the density matrices of the distilled state of a holographic quantum system. Of course, this definition depends on the specific distillation scheme, in other words, the specific model of our thread-state correspondence. However, these concise distillation schemes help us clarify the meaning and limitations of the PEE scheme. Finally, we will find that the PEE=CMI scheme cannot fully characterize the entanglement information of a quantum system, especially regarding the entanglement information of disconnected regions. We propose an improved approach to thinking about PEE, which involves studying the entanglement structure of a system with multi-partite entanglement, rather than the simple bipartite entanglement picture, to investigate how the various components of a chosen region $A$ (whether it is connected or disconnected) contribute to the von Neumann entropy of $A$ by entangling with other regions of the system.

The structure of this article is as follows: In section~\ref{sec2}, we reviewed the PEE=CMI scheme from two perspectives, which will reveal its close connection with bipartite entanglement. In section~\ref{sec3}, we first reviewed the concept of thread-state correspondence and applied it to explain PEE=CMI as bipartite entanglement, and present a definition of PEE based on the reduced density matrices of the holographic distilled state. Then, we analyzed the limitations of the bipartite entanglement interpretation when applied to disconnected regions, and pointed out a noteworthy coincidence between CMI and another quantum information theoretical quantity, tripartite information. In section~\ref{sec4}, we solved the limitations of the bipartite entanglement interpretation by introducing multipartite entanglement, which was achieved by introducing the extended $n$-thread/perfect tensor state correspondence. We also gave a definition of PEE based on the reduced density matrices of the holographic distilled state in the context of multipartite entanglement interpretation. Section~\ref{sec5} is a reflection and discussion on the concept of PEE, and discusses how to further develop the concept of thread-state correspondence when dividing the holographic quantum systems into more constituent parts. The final section is the conclusion and discussion.


\section{Review on the holographic PEE=CMI scheme}\label{sec2}


\subsection{The PEE manifestation}\label{sec21}
\begin{figure}[htbp]     \begin{center}
		\includegraphics[height=7.5cm,clip]{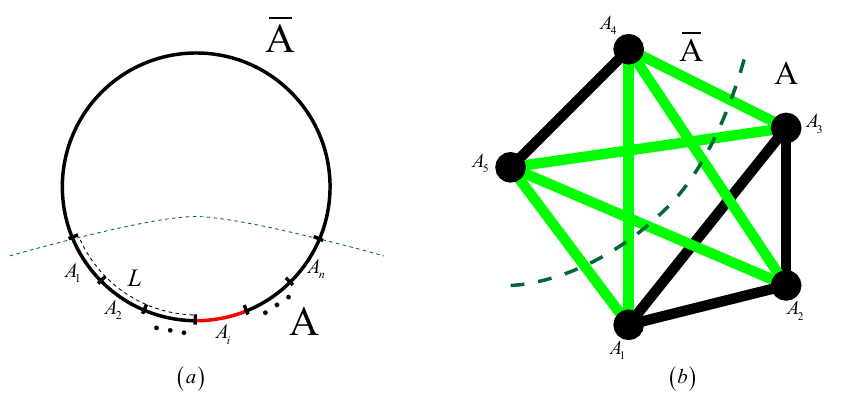}
		\caption{(a) (\ref{cmi}) represents the contribution of $A_i$ to the entanglement entropy $S(A)$ between $A$ and its complement $\bar{A}$.	(b) (\ref{equ}) represent that the entanglement entropy between $A$ and $\bar A$ comes from the sum of ${I_{ij}}$ connecting each elementary region ${A_i}$ inside $A$ and each elementary region ${A_j}$ inside the complement $\bar A$ (i.e., the green edges in the figure).}
		\label{fig21}
	\end{center}	
\end{figure}

Here we review the PEE=CMI scheme under the holographic framework~\cite{wen2020entanglement,Wen:2018whg,Kudler-Flam:2019oru} and its derivation based on the thread perspective~\cite{Lin:2021hqs}. To make the entire narrative more systematic, we start with such a setup (see figure~\ref{fig21}(a)): Consider a holographic quantum system in a pure state $\psi$ and one of its (connected) subsystems $A$. Decompose $A$ into several adjacent parts of very small size (let us assume that they are much larger than the Planck scale), i.e., $A = A_1 \cup A_2 \cup \cdots \cup A_n$~\footnote{For convenience, we agree that the ordering follows counterclockwise.}. Then we can define PEE $s_A(A_i)$ as the contribution of one of the small pieces $A_i$ to the entanglement entropy $S(A)$ between $A$ and its complement $\bar{A}$. Let $L = A_1 \cup A_2 \cup \cdots \cup A_{i-1}$ represent the region in front of $A_i$ (which also represents the spacing between $A_i$ and $\bar{A}$) in $A$, then the PEE=CMI scheme states that~\cite{Kudler-Flam:2019oru,Wen:2018whg,wen2020entanglement}
\begin{equation}\label{cmi}
	s_A(A_i) = \frac{1}{2} I(A_i,\bar{A} | L) \equiv \frac{1}{2} [S(A_i L) + S(\bar{A} L) - S(A_i \bar{A} L) - S(L)].
\end{equation}
Utilizing the property of the pure state, if we define $R = A_{i+1} \cup \cdots \cup A_n$, one can equivalently obtain
\begin{equation}\label{pee}
	s_A(A_i) = \frac{1}{2} [S(A_i L) + S(A_i R) - S(R) - S(L)],
\end{equation}
which is more commonly seen in the literature.

It is appropriate to add our first comment here. We believe that, so far, this characterization of quantum entanglement for $S(A)$ in PEE (or entanglement contour) manifestation, which focuses on the decompositon in terms of spatial components (or local points) within the $A$ region, is not the most natural. The limitation is reflected in the fact that the PEEs appear to depend on the artificially selected $A$. Given a specific region $A$, perhaps one would think that $s_A(A_i)$ characterizes the entanglement information of the quantum system in the small block $A_i$. The naive idea is that it can generally be considered as some kind of entanglement density $\rho(A_i)$. However, if we change the size of $A$ to $A'$, we will actually find that
\begin{equation}
	s_A(A_i) \neq s_{A'}(A_i),
\end{equation}
so the information of $s_A(A_i)$ in the original $A_i$ region seems very limited and can only be used to calculate $S(A)$. When we want to calculate $S(A')$, this information seems to be useless.

\subsection{The CMI manifestation}\label{subsec2.2}

Now we want to take another perspective to view the PEE=CMI scheme, especially in the holographic framework. We will focus on describing this in terms of pairs of points (or pairs of ``elementary regions") rather than the collection of local points (or the collection of ``elementary regions"). We can call the former perspective the PEE representation, and the perspective we are about to describe as the CMI representation. This CMI representation can be described in the formulation of kinematic space~\cite{Czech:2015kbp,Czech:2015qta} or in the formulation of a locking multiflow configuration in bit thread language~\cite{Lin:2021hqs}.

To do this, consider dividing the boundary quantum system into $N$ adjacent non-overlapping elementary regions, denoted as ${A_1}$, ${A_2}$, ..., ${A_N}$, where $N$ can be very large, so that the size of each elementary region is very small.~\footnote{However, for physical reasons, we will make sure that each small area is much larger than the Planck length so that the RT formula can be applied effectively.} A natural way to achieve this is to take $N$ elementary regions that are identical in size (which is easy to do for a one-dimensional space in $CFT_2$). Then, for each pair of elementary regions ${A_i}$ and ${A_j}$, we can define a function ${I_{ij}}$, which, in the context of~\cite{Lin:2021hqs}, is understood as the number of threads of the component flow ${{\vec v}_{ij}}$ of a locking bit thread configuration. We can describe this simply with a complete graph in graph theory (see figure~\ref{fig21}(b)), where each boundary vertex represents an elementary region and each edge is assigned the function ${I_{ij}}$. Inspired by the intuitive image of bit threads, \cite{Lin:2021hqs} proposed a scheme to depict the PEEs equivalently with a set $\left\{ {{I_{ij}}} \right\}$ containing a total of $N\left( {N - 1} \right)/2$ ${I_{ij}}$, where the ${I_{ij}}$ are defined such that:
\begin{equation}\label{equ} {S_{a\left( {a + 1} \right) \ldots b}} = \sum\limits_{i,j} {{I_{ij}}} ,\;\;\;{\rm{where}}\;i \in \left\{ {a,a + 1, \cdots ,b} \right\},\;j \notin \left\{ {a,a + 1, \cdots ,b} \right\}.\end{equation}
Here, ${S_{a(a+1)...b}}$ represents the entanglement entropy ${S_A}$ of a connected composite region ${A = A_{a(a+1)...b} \equiv A_a \cup A_{a+1} \cup \cdots \cup A_b}$. The equation can be intuitively understood as the entanglement entropy between $A$ and $\bar A$ comes from the sum of ${I_{ij}}$ connecting each elementary region ${A_i}$ inside $A$ and each elementary region ${A_j}$ inside the complement $\bar A$ (i.e., the green edges in figure~\ref{fig21}(b)). It can be verified that considering all possible ways of selecting the connected composite region, the system of equations~(\ref{equ}) includes a total of $N(N-1)/2$ constraints, so that the exact value of each ${I_{ij}}$ can be completely determined. Let ${\tilde L = A_{(i+1)...(j-1)}}$ be the region between ${A_i}$ and ${A_j}$ (which is a composite region consisting of many elementary regions and also represents the distance between ${A_i}$ and ${A_j}$), then the solution to ~(\ref{equ}) can be obtained as~\cite{Lin:2021hqs}:
\begin{equation}\label{iij}
	{I_{ij}} = \frac{1}{2}I({A_i},{A_j}\left| {\tilde L} \right.)
\end{equation}
Thus, we find that each function ${I_{ij}}$ is precisely given by the expression for CMI, which is why we have labeled it with the letter $I$. In fact, if we directly choose $\bar A = {A_j}$ in (\ref{iij}), we are led directly back to (\ref{pee}). In particular, when $L$ is the empty set, $I_{ij}$ is just half the mutual information between $A_i$ and $A_j$. Thus, given a specified connected region $A = {A_a} \cup {A_{a + 1}} \cup \cdots \cup {A_b}$, the thread formulation of the PEE contribution of the subregion $A_i$ to the entropy $S(A)$ is
\begin{equation}\label{add}{s_A}\left( {{A_i}} \right) = \sum\limits_j {{I_{ij}}} ,\;\;\;{\rm{where}}\;i \in \left\{ {a,a + 1, \cdots ,b} \right\},\;j \notin \left\{ {a,a + 1, \cdots ,b} \right\}
\end{equation}
By substituting the expression (\ref{iij}) and redefining $L = {A_a} \cup {A_{a + 1}} \cup \cdots \cup {A_{i - 1}}$, $R = {A_i} \cup {A_{i + 1}} \cup \cdots \cup {A_b}$, we can immediately obtain
\begin{equation}
	{s_A}({A_i}) = \frac{1}{2}I({A_i},\bar A\left| L \right.) = \frac{1}{2}(S({A_i}L) + S({A_i}R) - S(R) - S(L))
\end{equation}
which is completely consistent with (\ref{pee}).

We believe that from the perspective of CMI, that is, from the perspective of pairs of points rather than local points, it is more natural to consider the PEE scheme. In the previous section, we mentioned that ${s_A}\left( {{A_i}} \right)$ seems to only provide limited entanglement information that depends on the specific selection of $A$. However, from the perspective of CMI, ${s_A}\left( {{A_i}} \right)$ is actually ${I_{ij}}$, which can still provide useful data for the entanglement information of $S\left( {A'} \right)$ of another selected region $A'$. It's just that when calculating $S\left( {A'} \right)$, the ${I_{ij}}$ to be included are different (see (\ref{add})), but the data of $\left\{ {{I_{ij}}} \right\}$ is universal. In summary, the reason is that the perspective of pairs of points is more natural than that of local points when analyzing entanglement structure. In other words, kinematic space~\cite{Czech:2015kbp,Czech:2015qta} is more natural than original space when analyzing entanglement structure. A more detailed discussion of the connection between kinematic space and PEE=CMI scheme is presented in~\cite{Rolph:2021nan,Lin:2022flo}.


\section{The bipartite entanglement explanation of the PEE=CMI scheme }\label{sec3}

As we mentioned in the introduction, the fundamental definition of PEE based on the reduced density matrix has not yet been given. Our article is an attempt to construct a density matrix representation of PEE (at least within the holographic framework). This relies on the visualization tool of bit threads. \cite{Kudler-Flam:2019oru,Lin:2021hqs} first pointed out that the CMI representation of holographic PEE is very similar to the locking multiflow configuration of bit threads. In a locking bit thread configuration, each bit thread can be understood as corresponding to a specific (distilled) quantum state~\cite{Lin:2022agc,Lin:2022flo}. In this section, we will systematically restate these contents, leading to a density matrix representation of holographic PEE, at least in the sense of distilled states. However, we will also discuss the limitations of this distillation scheme and use an updated version to fix it later.

\subsection{Distilled density matrices of holographic PEE}\label{sec31}

\begin{figure}[htbp]     \begin{center}
		\includegraphics[height=7cm,clip]{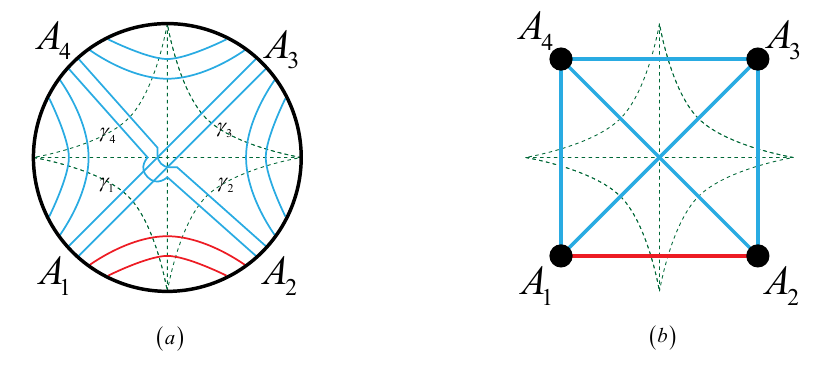}
		\caption{(a) A locking multiflow configuration in the simplified case of $N=4$, satisfying (\ref{equ2}). And in the sense of thread/state correspondence, this presents a special quantum state corresponding to (\ref{app}). (b) A simplified diagram corresponding to (a), where each edge now represents a thread bundle with capacity $F_{ij}$. And each edge explicitly corresponds to state (\ref{fstate}).}
		\label{fig32}
	\end{center}	
\end{figure}

To start with, note that in the context of bit threads, if we interpret $I_{ij}$, which is related to two elementary regions $A_i$ and $A_j$, as the number of threads in a locking multiflow configuration that only connects $A_i$ and $A_j$, then a corresponding locking bit thread configuration can be used to characterize the entanglement structure containing the information  of $\left\{ {{I_{ij}}} \right\}$~\cite{Lin:2021hqs}. A simple explanation: multiflow (or multi-commodity flow) is a term analogous to network flow theory. Dividing the boundary quantum system into $N$ adjacent and non-overlapping elementary regions, the multiflow is defined as the set of a total of $N(N-1)/2$ component flows ${v_{ij}}$, each describing a bundle of uninterrupted threads whose endpoints are constrained on $A_i$ and $A_j$ respectively.~\footnote{ Note that for the purposes of this discussion, these threads do not need to be strictly understood as bit threads, which require more non-trivial constraints on the density of threads in the bulk. More detailed discussions can be found in~\cite{Lin:2021hqs}.} A locking multiflow configuration refers to such a configuration in which the numbers of threads passing through the RT surfaces corresponding to a given set of boundary subregions are exactly equal to the areas of these RT surfaces (up to a factor of $4G$), and thus equal to the entanglement entropies of the boundary subregions by RT formula (see figure~\ref{fig32}(a)).

Let's explain how to describe the PEE=CMI scheme using a locking bit thread configuration in the simplified case of $N=4$, where there are only four elementary regions. As shown in figure~\ref{fig32}(a), we explicitly draw all threads in each component flow $v_{ij}$ of the locking multiflow (this oversimplified figure assumes that all $F_{ij}$ are equal to 2). Let $F_{ij}$ (named component flow flux, CFF) denotes the number of threads in the bundle $v_{ij}$, then we can construct a locking multiflow configuration where $\left\{ {{F_{ij}}} \right\}$ satisfies:
\begin{equation}\label{equ2}
	S_{a(a+1)\cdots b} = \sum_{i,j} F_{ij}, \quad \text{where } i \in {a,a+1,\cdots,b}; j \notin {a,a+1,\cdots,b}.
\end{equation}
Obviously, this is consistent with (\ref{equ}), and we obtain a one-to-one correspondence between $I_{ij}$ and $F_{ij}$. In this model, the CMI (or PEE) can be simply understood as the flux of bit thread bundles, i.e.,
\begin{equation}\label{fij}
	I_{ij} = F_{ij}.
\end{equation}
Furthermore, for a connected region $A = A_a \cup A_{a+1} \cup \cdots \cup A_b$, whose von Neumann entropy is given by (\ref{equ2}), the PEE contribution of a component $A_i$ to the entropy $S(A)$ can be expressed in the thread formulation as:
\begin{equation}\label{add}
	s_A(A_i) = \sum_j F_{ij}, \quad \text{where } i \in {a,a+1,\cdots,b}; j \notin {a,a+1,\cdots,b}.
\end{equation}

A more explicit physical interpretation is crucial. Inspired by the study of the corresponding states of RT surfaces in the surface/states correspondence~\cite{Miyaji:2015fia,Miyaji:2015yva},~\cite{Lin:2022flo} proposed the instructive
thread/state correspondence, which assumes that in a locking multiflow configuration, each thread corresponds to a quantum superposition state of $J$ orthogonal states with equal probabilities. In particular, we can assume that $J = 2$ and label the two orthogonal states as the red state $\left|r\right\rangle$ and the blue state $\left|b\right\rangle$, i.e.,
\begin{equation}\label{thr}
	\left|\text{thread}\right\rangle = \frac{1}{\sqrt{2}} \left(\left|r\right\rangle + \left|b\right\rangle\right).
\end{equation}
In the context of the surface/state correspondence, the thread state~(\ref{thr}) provide the quantum states corresponding to the RT surfaces of all elementary regions and connected composite regions, as well as their entanglement structures with each other. For example, as shown in figure~\ref{fig32}(a), consider the closed surface  $\Gamma = \gamma_1 \cup \gamma_2 \cup \gamma_3 \cup \gamma_4$ consisting of the four RT surfaces corresponding to all elementary regions $A_1$, $A_2$, $A_3$, $A_4$. According to the surface/state correspondence, it should correspond to a pure state $\left| \Psi_{\Gamma} \right\rangle$, which can be expressed as a quantum superposition formally:
\begin{equation}\label{psi}
	\left| \Psi_{\Gamma} \right\rangle = \sum_Q C_Q \left| Q \right\rangle, \quad \text{where} \quad \left| Q \right\rangle \equiv \left| \gamma_1 \right\rangle \otimes \left| \gamma_2 \right\rangle \otimes \left| \gamma_3 \right\rangle \otimes \left| \gamma_4 \right\rangle.
\end{equation}
\cite{Lin:2022flo} proposed the thread/state rules, which state that: (1) The red state of the thread corresponds to a basis state $\left| 0 \right\rangle$ in the surface/state context, while the blue state corresponds to a basis state $\left| 1 \right\rangle$. (2) When measuring, the same thread is always in the same color state, and different threads are simply assumed not to affect each other. Then according to this prescription, the states of a series of RT surfaces in the surface/state correspondence can be given by the state of a locking thread configuration. Let us show in principle how to use the thread/state rules to give the state of $\left| \Psi_{\Gamma} \right\rangle$. Suppose we measure the probability amplitudes for the thread configurations to be in various color states. Figure~\ref{fig32}(a) shows a special case where we measure the thread configuration and find that all threads connecting $A_1$ to $A_2$ are in the red state, while all other threads are in the blue state. Then, the $\Gamma$ surface is in such a state:
\begin{equation}\label{app}
	\left| Q \right\rangle = \left| 111100 \right\rangle \otimes \left| 001111 \right\rangle \otimes \left| 111111 \right\rangle \otimes \left| 111111 \right\rangle
\end{equation}
And we can obtain from the thread/state rules that
\begin{equation}
	C_Q = \left( \frac{1}{\sqrt{2}} \right)^{\sum F_{ij}} = \left( \frac{1}{\sqrt{2}} \right)^{12}
\end{equation}
The situation could also be described as follows: for a specified closed surface, each thread corresponds to a state expressed as
\begin{equation}\label{bell}
	\left| {{\rm{thread}}} \right\rangle = \frac{1}{{\sqrt 2 }}\left( {\left| {{\rm{00}}} \right\rangle + \left| {{\rm{11}}} \right\rangle } \right),
\end{equation}
where the two qubits correspond to the ingoing and outgoing intersections of the thread with the closed surface respectively.\footnote{This interpretation of thread state also appears in~\cite{Harper:2022sky}, although in a slightly different form.} Therefore, a thread can also be understood as corresponding to a Bell pair. Anyway, the same method can be used to completely write out each term in (\ref{psi}), thus fully constructing the specific expression of $\left| {{\Psi _\Gamma }} \right\rangle$. In the simplified example presented here, the entire holographic quantum system is divided into four parts, and $\left| {{\Psi _\Gamma }} \right\rangle$ can be considered as the distilled state of the entire quantum system at this level, while the density matrix corresponding to each RT surface corresponds to the distilled density matrix of its corresponding boundary subregion~\cite{lin2020surface,lin2021bit,bao2019beyond,Bao:2019fpq}. It is not difficult to directly generalize this procedure to the case of large $N$, as in the previous subsection, and then we will obtain a more refined distilled state.

Within this framework of understanding a holographic quantum system with a distilled state, it is straightforward to write the density matrix representation of PEE. The simplified diagram as shown in figure~\ref{fig32}(b) is still useful, where each edge now represents a thread bundle with capacity $F_{ij}$. By (\ref{thr}) and (\ref{bell}), each edge explicitly corresponds to a state, which is the direct product state of all the states of the threads contained in the bundle:
\begin{equation}\label{fstate}
	\left| {{F_{ij}}} \right\rangle = {(\frac{1}{{\sqrt 2 }}(\left| r \right\rangle + \left| b \right\rangle ))^{ \otimes {F_{ij}}}} = {(\frac{1}{{\sqrt 2 }}({\left| 0 \right\rangle _i} \otimes {\left| 0 \right\rangle _j} + {\left| 1 \right\rangle _i} \otimes {\left| 1 \right\rangle _j}))^{ \otimes {F_{ij}}}},
\end{equation}
where, for example, ${\left| 0 \right\rangle _i}$ represents that a ``distilled qubit'' in the elementary region ${A_i}$ is in the $0$ state indicated by thread/state rules, etc. Alternatively, (\ref{fstate}) can be written as
\begin{equation}
	\left| {{F_{ij}}} \right\rangle = \sum\limits_{{q_{ij}} = 0}^{{2^{^{{F_{ij}}}}} - 1} {{{(\frac{1}{{\sqrt 2 }})}^{{F_{ij}}}}} \left| {{q_{ij}}} \right\rangle,
\end{equation}
where each ${\left| {{q_{ij}}} \right\rangle }$ represents a basis state of the overall configuration of all the threads in the bundle ${v_{ij}}$, and there are a total of ${2^{{F_{ij}}}}$ such states. Now, by thread/state rules and (\ref{equ2}), we can write the distillated state of the entire quantum system, which is the direct product of the states of all the threads corresponding to the locking thread configuration, or equivalently, the direct product of the states of all the edges in the complete graph~\ref{fig32}(b), given by:
\begin{equation}\label{psi}
	\left| {{\Psi _\Gamma }} \right\rangle  = \mathop  \otimes \limits_{{\rm{all}}\;ij} \left| {{F_{ij}}} \right\rangle  = \mathop  \otimes \limits_{{\rm{all}}\;ij} {(\frac{1}{{\sqrt 2 }}({\left| 0 \right\rangle _i} \otimes {\left| 0 \right\rangle _j} + {\left| 1 \right\rangle _i} \otimes {\left| 1 \right\rangle _j}))^{ \otimes {F_{ij}}}}
\end{equation}
On the other hand, for a subregion $A = A_a \cup A_{a+1} \cup \cdots \cup A_b$, its reduced density matrix is given by:
\begin{equation}
	\rho_A = tr_{\bar{A}} \left| \Psi_{\Gamma} \right\rangle \left\langle \Psi_{\Gamma} \right|
\end{equation}
Substituting in (\ref{psi}), we get:
\begin{equation}
	\rho_A = \bigotimes_{ij} tr_{A_j} \left| F_{ij} \right\rangle \left\langle F_{ij} \right| = \bigotimes_{ij} \left( \frac{1}{2} \left| 0 \right\rangle_i \left\langle 0 \right|_i + \frac{1}{2} \left| 1 \right\rangle_i \left\langle 1 \right|_i \right)^{\otimes F_{ij}} ,
\end{equation}
where $i \in \left\{ {a,a + 1, \cdots ,b} \right\},\;j \notin \left\{ {a,a + 1, \cdots ,b} \right\}$.
Clearly, from (\ref{equ2}), we can verify that:
\begin{equation}\label{sa}
	-{\mathrm{tr}}(\rho_A \log_2 \rho_A) = S_A
\end{equation}

Now, the expressions for CMI or PEE based on the density matrix are about to be revealed. Note that the operation $-\mathrm{tr} (\rho_A \log \rho_A)$ essentially counts the number of threads the product state of which gives the correct density matrix $\rho_A$, since this number gives the von Neumann entropy $S(A)$ of $A$. Inspired by this, we realize that counting the number of threads in the thread bundle ${v_{ij}}$ actually gives the CMI contribution to $S(A)$. Therefore, we can first define a quantity called ``the density matrix of CMI":
\begin{equation}
	\rho_{ij} = \bigotimes tr_{A_j} \left| F_{ij} \right\rangle \left\langle F_{ij} \right| = \bigotimes \left( \frac{1}{2} \left| 0 \right\rangle_i \left\langle 0 \right|_i + \frac{1}{2} \left| 1 \right\rangle_i \left\langle 1 \right|_i \right)^{\otimes F{_{ij}}}
\end{equation}
Then, following (\ref{sa}), we can define the density matrix-based definition of CMI as:
\begin{equation}
	I_{ij} = -\mathrm{tr}(\rho_{ij} \log_2 \rho_{ij})
\end{equation}
In addition, if we choose the PEE representation, for a component $A_i$ in $A = A_a \cup A_{a+1} \cup \cdots \cup A_b$, by (\ref{add}) and the same logic, we can define a quantity called ``the reduced density matrix of PEE":
\begin{equation}
	{\rho_{{A_i} \to A}} = \mathop  \otimes \limits_j t{r_{{A_j}}}\left| {{F_{ij}}} \right\rangle \left\langle {{F_{ij}}} \right| = \mathop  \otimes \limits_j {(\frac{1}{2}{\left| 0 \right\rangle _i}{\left\langle 0 \right|_i} + \frac{1}{2}{\left| 1 \right\rangle _i}{\left\langle 1 \right|_i})^{ \otimes {F_{ij}}}},
\end{equation}
where $i \in \left\{ {a,a + 1, \cdots ,b} \right\},\;j \notin \left\{ {a,a + 1, \cdots ,b} \right\}$. Then we obtain the expression of PEE based on the density matrix:
\begin{equation}
	{s_A}({A_i}) =  - tr({\rho _{{A_i} \to A}}{\log _2}{\rho _{{A_i} \to A}}),
\end{equation}
which will indeed go back to (\ref{add}).


\subsection{Limitations of the distilled bipartite entanglement interpretation}\label{sec32}
In some sense the materials reviewed in the previous section are sufficient to convince us that PEE or CMI should be understood as distilled bipartite entanglement in the sense of distillation, i.e., one should not expect it to faithfully reflect the complete entanglement structure of a quantum system. However, this section will point out two issues related to PEE. The first will show that even in the sense of entanglement distillation, imagining more complex multipartite entanglement is necessary to characterize more information about the entanglement structure of the quantum system. The second will point out an interesting coincidence between CMI and tripartite information (TI) in holographic duality. These phenomena will be resolved in the section~\ref{sec4} by introducing multipartite entanglement.
\subsubsection{A paradox related to disconnected regions}\label{sec321}

Here, we present the limitations of using distilled bipartite entanglement to understand the PEE=CMI scheme. We will demonstrate this with a crucial thought experiment. We still use the example of the four-partite graph from the previous subsection, but now we consider the case where the sizes of $A_1$ and $A_3$ are relatively small (see figure~\ref{fig33}(b) and figure~\ref{fig32}(a)). Then, we consider a disconnected region $B={A_1}\cup{A_3}$, so we should have 
\begin{equation}\label{exp}
S(B)=S({A_1})+S({A_3}). 
\end{equation}
Here, a paradox arises. The point is that, according to the thread-state correspondence, the threads connecting $A_1$ and $A_3$ should be considered as internal entanglement of $B={A_1}\cup{A_3}$, so they cannot contribute to $S(B)$.

Based on experience from the locking bit thread configuration, we have the following argument: In figure~\ref{fig33}(b), the entanglement entropy of $A_1$ should be equal to the number of threads crossing the corresponding RT surface ${\gamma _1}$ (which equals the area of ${\gamma _1}$), and similarly the entanglement entropy of $A_3$ should be equal to the number of threads crossing the corresponding RT surface ${\gamma _3}$ (which equals the area of ${\gamma _3}$). However, according to the RT formula, mathematically we expect the entropy of $B={A_1}\cup{A_3}$ to be equal to the sum of the areas of ${\gamma _1}$ and ${\gamma _3}$, which should be equal to the number of threads crossing these two surfaces, including twice the number of threads connecting $A_1$ and $A_3$. But according to the thread-state interpretation, these threads connecting $A_1$ and $A_3$ represent internal entanglement of $B$, in other words, they only distill the entanglement inside $B$ into Bell pairs, rather than distilling the entanglement between $B$ and its complement into Bell pairs, so they cannot contribute to $S(B)$. Let us further explicitly clarify this from the distilled state (\ref{psi}) of the system. It can be verified that it will calculate 
\begin{equation}
	S(B)={F_{14}}+{F_{24}}+{F_{23}}+{F_{34}}<S({A_1})+S({A_3})={F_{14}}+{F_{24}}+{F_{23}}+{F_{34}}+2{F_{13}},
\end{equation} 
which is inconsistent with our expectation~(\ref{exp}). The reason is that, according to the thread-state viewpoint, the threads connecting $A_1$ and $A_3$ represent internal entanglement of $B$ and cannot be included in the entanglement entropy of $B$ and its complement. We thus find that only bipartite entanglement cannot give the correct data for $S({A_1})$, $S({A_3})$, and $S\left( {{A_1} \cup {A_3}} \right)$ simultaneously. Using bipartite entanglement to characterize the entanglement information of the PEE=CMI scheme has limitations. 

Therefore, this leads us to the following comments: Conclusion 1: We must consider more complex multipartite entanglement. Conclusion 2: Since the entire entanglement structure of a quantum system cannot be described, the bipartite entanglement interpretation of the PEE=CMI scheme should be merely understood in the following sense: by approximating (reorganizing and mapping) the quantum state of each elementary region as some ``distilled states"~\cite{lin2020surface,lin2021bit,bao2019beyond,Bao:2019fpq}, at least the entanglement entropy for the connected regions can be characterized correctly.

\subsubsection{A coincidence between CMI and tripartite information}\label{sec322}
\begin{figure}[htbp]     \begin{center}
		\includegraphics[height=7cm,clip]{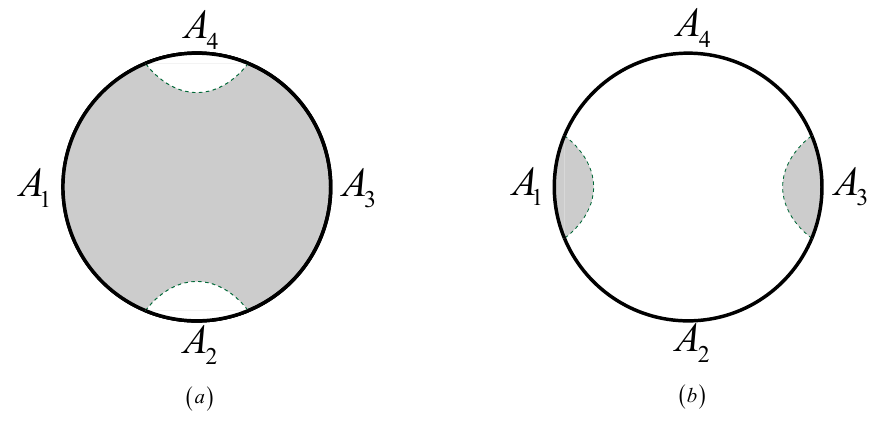}
		\caption{(a) When the sizes of $A_2$ and $A_4$ are relatively small, we have (\ref{24}). (b) When the sizes of $A_1$ and $A_3$ are relatively small, we have (\ref{13}). The grey shaded area represents the entanglement wedge of ${{A_1} \cup {A_3}}$ in both cases.}
		\label{fig33}
	\end{center}	
\end{figure}
This subsection will point out a noteworthy coincidence between CMI and tripartite information (TI) in the framework of the holographic principle. In the next subsection, we will explain why this coincidence occurs, which is actually related to the necessary appearance of multipartite entanglement.

Tripartite information (TI) is also a fundamental quantity in (quantum) information theory, similar to CMI, and it characterizes correlations involving three systems. It has also been studied in the context of holographic duality (see e.g. \cite{Cui:2018dyq,rangamani2017holographic,alishahiha2015time,mozaffar2015holographic,hayden2013holographic,asadi2018holographic,mahapatra2019interplay,mirabi2016monogamy,Ju:2023tvo}). Given three subsystems $A_1, A_2,$ and $A_3$, TI is defined as
\begin{equation}
	I({A_1}:{A_2}:{A_3}) = S({A_1}{A_2}) + S({A_2}{A_3}) + S({A_1}{A_3}) - S({A_1}) - S({A_2}) - S({A_3}) - S({A_1}{A_2}{A_3}),
\end{equation}
where $\overline{{A_1}{A_2}{A_3}} = {A_4}$ is defined. It can be verified that TI is symmetric under permutations of $A_1, A_2, A_3,$ and $A_4$, and it will be seen later that this is related to the fact that it is a kind of four-partite entanglement~\cite{Cui:2018dyq,Harper:2022sky}.~\footnote{ Moreover, for convenience, the TI we define here differs from the one defined in traditional literature by a minus sign.}

Now, in the framework of the holographic principle, let us continue to consider a pure state quantum system divided into four parts. More explicitly, consider two cases as shown in figure~\ref{fig33}. In the first case, since the sizes of $A_2$ and $A_4$ are relatively small, we have
\begin{equation}\label{24}
	S({A_1}{A_3}) = S({A_2}) + S({A_4}),
\end{equation}
which, when substituted into the definition of TI, leads to a coincidence:
\begin{equation}\label{i1}
	I({A_1}:{A_2}:{A_3}) = I({A_2}:{A_4}|{A_1}) = 2{F_{24}},
\end{equation}
where $F_{24}$ denotes the entanglement between $A_2$ and $A_4$. On the other hand, for the case where $A_1$ and $A_3$ are relatively small, we have
\begin{equation}\label{13}
	S({A_1}{A_3}) = S({A_1}) + S({A_3}),
\end{equation}
which leads to
\begin{equation}
	I({A_1}:{A_2}:{A_3}) = I({A_1}:{A_3}|{A_2}) = 2{F_{13}}.
\end{equation}
This is an interesting coincidence that should have an explanation. In the next subsection, we will provide an explanation by considering distilled states involving multipartite entanglement.

\section{The multipartite entanglement explanation of the PEE=CMI scheme }\label{sec4}


The previous discussion has made us realize that introducing multipartite entanglement to construct finer distilled states is a possible solution to overcome limitations of the correct characterization of the entropies for disconnected regions. Here we provide an explicit scheme. The materials have already been prepared in the literature~\cite{He:2019ttu}\cite{Harper:2022sky}. \cite{He:2019ttu} proposes the K-basis method of holographic entropy cone theory, and \cite{Harper:2022sky} proposes the so-called hyperthread/perfect state correspondence (also seen in earlier discussions in~\cite{Harper:2021uuq,Cui:2018dyq}), which can be regarded as a generalization of the thread/state correspondence reviewed in section~\ref{sec31}, although it is expressed in a slightly different way.

\subsection{$n$-thread/perfect tensor state correspondence}
\begin{figure}[htbp]     \begin{center}
\includegraphics[height=8cm,clip]{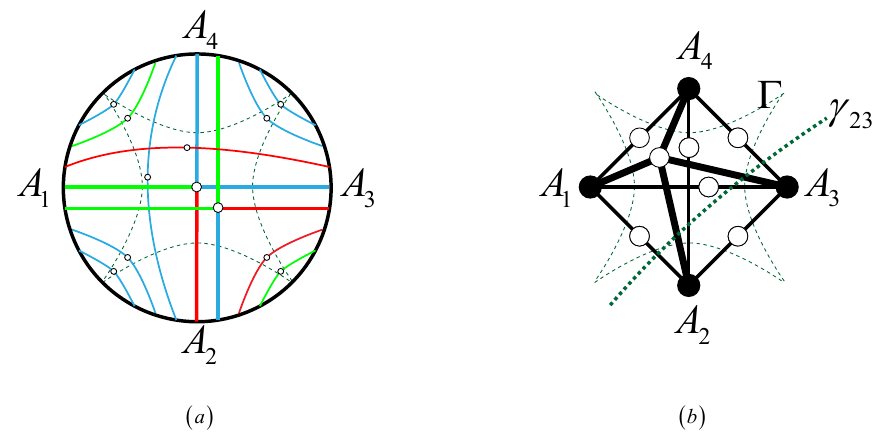}
		\caption{(a) Schematic diagram of a locking $n$-thread configuration and the generalized $n$-thread/perfect tensor state correspondence. Each  $n$-thread have $n$ legs extending from an internal vertex. The original bit threads can be seen as 2-threads.	(b) The simplified diagram of (a), where the elementary regions are depicted as the boundary vertex (marked with black dots), and an ``$n$-thread bundle" is simply described by $n$ legs extending from an $n$-valent internal vertex (marked with an empty dot) connected to the boundary vertex, with each leg assigned the value of the number of $n$-threads.	}
		\label{fig41}
	\end{center}	
\end{figure}

In simple terms, an $n$-hyperthread is a generalization of a bit thread~\cite{Harper:2021uuq,Harper:2022sky}. A bit thread has two endpoints, while an $n$-hyperthread is defined as having $n$ endpoints extending from an internal vertex, and these endpoints are tethered to the holographic boundary. One can also add an auxiliary internal vertex in the middle of the original bit thread and coordinate it as a $2$-hyperthread. We will hereafter refer to it as an $n$-thread  for convenience. It is proposed that an $n$-thread can be associated with a perfect tensor state~\cite{Harper:2022sky}. An $n$-perfect tensor state $\left| {P{T_n}} \right\rangle$ is defined as a pure state of $2n$ spins with a special property: any set of $n$ spins has maximal entanglement with the complementary set of $n$ spins. Perfect tensor states are the important ingredients for quantum error-correcting codes in the famous holographic HaPPY code~\cite{pastawski2015holographic}. They are also known as absolutely maximally entangled (AME) states in quantum information theory~\cite{helwig2012absolute,helwig2013absolutely}.

We will restate these things in our language to fit the purpose of revising the bipartite entanglement distillation description of the PEE=CMI scheme. As shown in figure~\ref{fig41}(a), to solve the difficulty in section~\ref{sec3}, we not only introduce 2-threads (note that we now draw an auxiliary internal vertex for it in the figure), but also introduce 4-threads. Intuitively speaking, similar to what one did in the holographic tensor network model (we will explain this from the perspective of thread-state later): the entanglement entropy of the boundary subregion (or equivalently, the area of its corresponding RT surface) is given by the number of legs cut off by its corresponding RT surface. Or, in terms of bit threads, entropy is given by the number of threads passing through the RT surface. Let the number of 2-threads, i.e., the number of 2-legged internal vertices, be denoted as $F$ (consistent with the convention in section~\ref{sec2}), and let the number of 4-threads, i.e., the number of 4-legged internal vertices, be denoted as $K$ (consistent with the convention in~\cite{Harper:2022sky}), and use subscripts to represent the boundary elementary regions connected by $n$-threads. Then we obtain the following equation set in the locking $n$-thread configuration~\cite{Harper:2022sky,He:2019ttu}:
\begin{equation}\label{set}\begin{array}{l}
	S({A_1}) = {F_{12}} + {F_{13}} + {F_{14}} + {K_{1234}}\\
	S({A_2}) = {F_{12}} + {F_{23}} + {F_{24}} + {K_{1234}}\\
	S({A_3}) = {F_{13}} + {F_{23}} + {F_{34}} + {K_{1234}}\\
	S({A_1}{A_2}{A_3}) = {F_{14}} + {F_{24}} + {F_{34}} + {K_{1234}} = S({A_4})\\
	S({A_1}{A_2}) = {F_{13}} + {F_{14}} + {F_{23}} + {F_{24}} + 2{K_{1234}} = S({A_3}{A_4})\\
	S({A_2}{A_3}) = {F_{12}} + {F_{24}} + {F_{13}} + {F_{34}} + 2{K_{1234}} = S({A_1}{A_4})\\
	S({A_1}{A_3}) = {F_{12}} + {F_{14}} + {F_{23}} + {F_{34}} + 2{K_{1234}} = S({A_2}{A_4})
\end{array}.\end{equation}
This set of equations has 7 constraints about entropies, which we can represent as:
\begin{equation}
{S^3} = \{ S({A_1}),S({A_2}),S({A_3}),S({A_1}{A_2}),S({A_2}{A_3}),S({A_1}{A_3}),S({A_1}{A_2}{A_3})\} .
\end{equation}
In the language of the holographic entropy cone, this is called a 7-dimensional (given by ${7 = 2^3-1}$) entropy vector corresponding to the three boundary regions $A_1$, $A_2$, $A_3$, and a ``puriﬁer'' $A_4$~\cite{Bao:2015bfa}.
At the same time, we have 7 unknowns represented as:
\begin{equation}
{K^3} = \{ {F_{12}},{F_{13}},{F_{14}},{F_{23}},{F_{24}},{F_{34}},{K_{1234}}\} ,
\end{equation}
which is called the K-basis of the entropy vector in~\cite{He:2019ttu}. We can then solve this equation set for a unique solution. Interestingly, we obtain:
\begin{equation}\label{sol}
	{F_{ij}} = \frac{1}{2}(S({A_i}) + S({A_j}) - S({A_i}{A_j})) \equiv \frac{1}{2}I({A_i}:{A_j}),
\end{equation}
\begin{equation}\label{k1234}
	{K_{1234}} = \frac{1}{2}(S({A_1}{A_2}) + S({A_2}{A_3}) + S({A_1}{A_3}) - S({A_1}) - S({A_2}) - S({A_3}) - S({A_1}{A_2}{A_3})) \equiv \frac{1}{2}I({A_1}:{A_2}:{A_3}).
\end{equation}
The solution (\ref{k1234}) is worth noting, as it suggests that the revision of the PEE=CMI scheme using 4-threads may also provide an explanation for the coincidence between CMI and TM in the holographic context discussed in Section~\ref{sec32}. In particular, from the last equation in the (\ref{set}), we see that after introducing the 4-threads, the locking thread configuration can correctly give the entanglement entropies for the disconnected regions ${{A_1} \cup {A_3}}$ and ${{A_2} \cup {A_4}}$. In any case, it is important to note that even in the current situation, these quantities should be understood in terms of entanglement distillation, and should not be expected to provide a complete picture of entanglement structure.

Let us now clarify this distilled state from the perspective of thread-state correspondence. To do this, we need to update the thread/state rules reviewed in Section~\ref{sec31} to accommodate the current situation. Since each thread is now viewed as a 2-thread with an internal vertex, we rewrite the rule~(\ref{thr}) as:
\begin{equation}
	\left| {2 - thread} \right\rangle = \frac{1}{{\sqrt 2 }}\left( {\left| {rr} \right\rangle + \left| {bb} \right\rangle } \right)
,\end{equation}
which represents that when one of the legs of the 2-thread is in the red state, the other leg must also be in the red state, and similarly, when one leg is in the blue state, the other leg must also be in the blue state. In fact, in this section, we adopt another convention, namely, we assign three colors to the states of a 2-thread, namely red, blue, and green. This does not make a substantive difference, but it causes the logarithmic base to change from 2 to 3 when computing entropies. Therefore, in this section, we adopt the following states:
\begin{equation}\label{2thr}
	\left| {2 - thread} \right\rangle = \frac{1}{{\sqrt 3 }}\left( {\left| {rr} \right\rangle + \left| {bb} \right\rangle + \left| {gg} \right\rangle } \right)
.\end{equation}
In other words, this now represents a ``qutrit thread". On the other hand, as pointed out by~\cite{Harper:2022sky}, actually, a general $n$-thread corresponds to an $n$-perfect tensor state. In particular, a special 4-perfect tensor state has been explicitly constructed in~\cite{pastawski2015holographic}. In the context of quantum error-correcting codes, it is also known as a 3-qutrit code. Therefore, we propose a 4-thread corresponds to the following 3-qutrit code state:~\footnote{Let us define the ordering follows the counterclockwise direction.}
\begin{equation}\label{4thr}
\begin{array}{l}
	\left| {4 - thread} \right\rangle  = \frac{1}{3}(\left| {rrrr} \right\rangle  + \left| {bbbr} \right\rangle  + \left| {gggr} \right\rangle \\
	\quad \quad \quad \quad \quad \quad  + \left| {rbgb} \right\rangle  + \left| {bgrb} \right\rangle  + \left| {grbb} \right\rangle \\
	\quad \quad \quad \quad \quad \quad  + \left| {rgbg} \right\rangle  + \left| {brgg} \right\rangle  + \left| {gbrg} \right\rangle )
\end{array}
.\end{equation}
Similarly, we extend the thread/state rules to the case of $n$-threads. An $n$-thread has $n$ legs, but the states of each leg are entangled with each other in the manner of (\ref{4thr}). And similarly, in the sense of surface/state correspondence or holographic distillation, the red, blue, and green states correspond to $\left| 0 \right\rangle ,\;\left| 1 \right\rangle \;{\rm{and}}\;\left| 2 \right\rangle$ states, respectively. More specifically, in the context of surface-state correspondence, when we consider the state of an RT surface ${\gamma _R}$ corresponding to a boundary subregion $R$ (which can now be connected or disconnected), legs in the red, blue, and green states respectively indicate the distilled states that can be represented by $\left| 0 \right\rangle ,\;\left| 1 \right\rangle \;{\rm{and}}\;\left| 2 \right\rangle$. For convenience, in the following discussion, we will no longer distinguish between red, blue, and green and $\left| 0 \right\rangle ,\;\left| 1 \right\rangle \;{\rm{and}}\;\left| 2 \right\rangle$. Anyway, the logic of section~\ref{sec31} can be directly extended to this case.

To see how this thread-state scheme can provide entanglement distillation schemes beyond the version of bipartite entanglement, that is, correctly provide the entanglement entropies of connected and disconnected subregions that satisfy equation (\ref{equ}), we first note that for each 4-thread, the entanglement entropy between any single leg and the other three legs is given by $ln3$, while the entanglement entropy between any set of two legs and their complements is given by $2ln3$. In fact, this is a general property of perfect tensor states. On the other hand, in the characterization of the state corresponding to 2-threads (\ref{2thr}), which can also be regarded as a perfect tensor state, the entanglement entropy between any single leg and the remaining leg is also given by $ln3$. Therefore, as mentioned earlier, we can now formulate the calculation of entanglement entropy in a way similar to that done in holographic tensor network models: as shown in figure~\ref{fig41}, the entanglement entropy of a (connected or disconnected) boundary subregion $R$ is calculated by cutting the entire locking thread configuration into two halves using a cut, where one part is completely adjacent to $R$ and the other part is completely adjacent to its complement $\bar R$. The number of cut legs multiplied by the constant $ln3$ gives the correct entanglement entropy:
\begin{equation}
	\text{entanglement entropy of a (connected or disconnected) boundary subregion} = \text{number of cut legs} \times \ln 3
\end{equation}

\subsection{Distilled density matrices of holographic PEE: an improved version}\label{sec42}
\begin{figure}[htbp]     \begin{center}
		\includegraphics[height=10cm,clip]{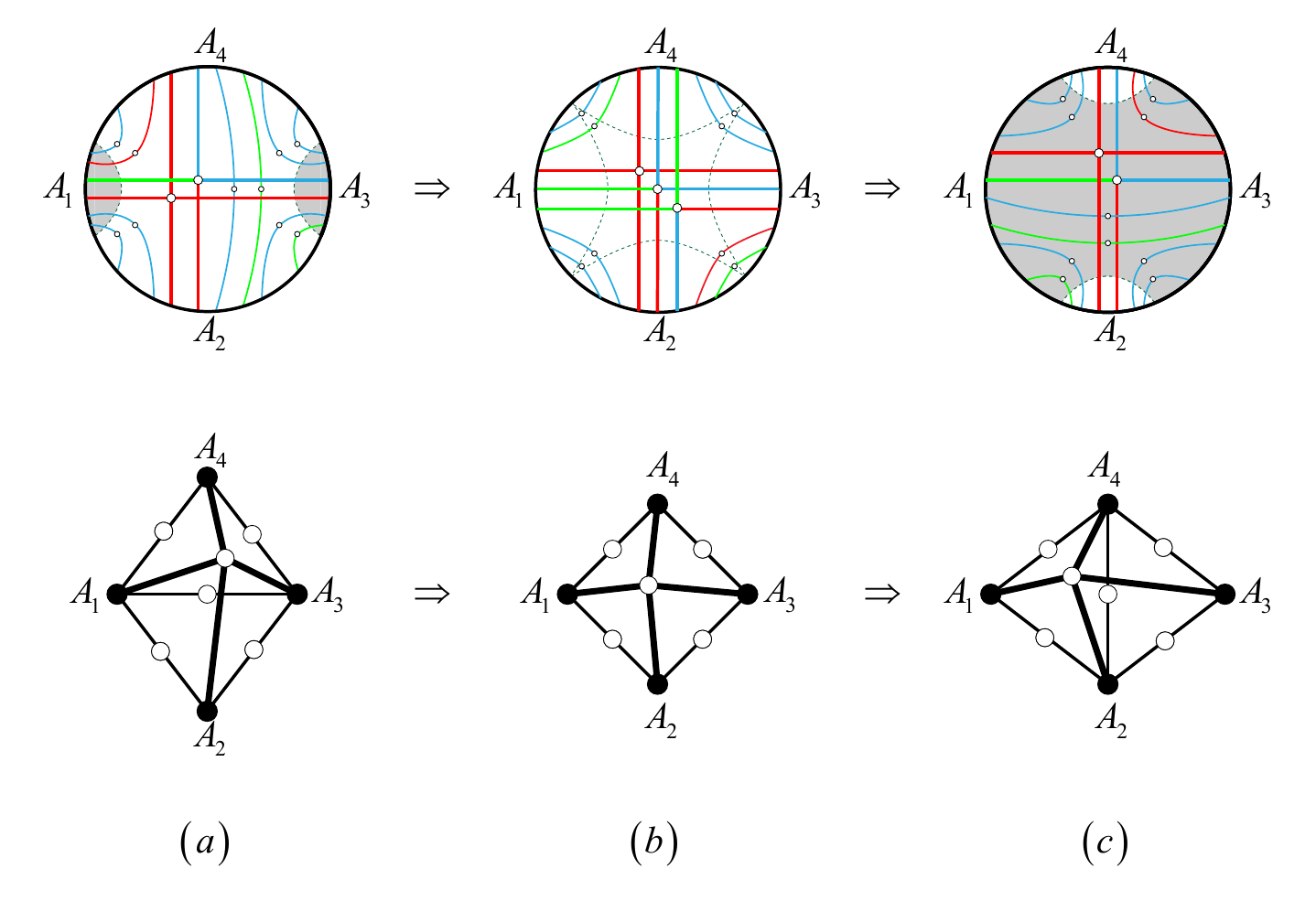}
		\caption{For different four-partition cases of the quantum system, some thread bundles can disappear, and if we continuously adjust the way of partition, a phase transition will occur. The upper part of the figure shows the change of the locking $n$-thread configuration, and the lower part shows the corresponding simplified diagram.}
		\label{fig42}
	\end{center}	
\end{figure}

Let us address in more detail the issue raised in section~\ref{sec3}. As shown in figure~\ref{fig42}, the analysis will involve a phase transition process. We still take the example of dividing the quantum system into four parts, and the case of dividing it into more parts will be discussed in the next section. First, we consider the case in figure~\ref{fig42}(a), where the sizes of $A_1$ and $A_3$ are relatively small, so the entanglement entropy between ${A_1} \cup {A_3}$ and ${A_2} \cup {A_4}$ is given by the area of the RT surface of $A_1$ and $A_3$. At this point, according to Eq.~(\ref{sol}), we find that there is a special property given by
 \begin{equation}\label
 	{f13}{F_{13}}=0,
\end{equation}
which allows us to overcome the two problems discussed in section ~\ref{sec3}. To see this, we write explicitly the form of the (\ref{set}) induced by (\ref{f13}):
\begin{equation}\label{set1}\begin{array}{l}
	S({A_1}) = {F_{12}} + {F_{14}} + {K_{1234}}\\
	S({A_2}) = {F_{12}} + {F_{23}} + {F_{24}} + {K_{1234}}\\
	S({A_3}) = {F_{23}} + {F_{34}} + {K_{1234}}\\
	S({A_1}{A_2}{A_3}) = {F_{14}} + {F_{24}} + {F_{34}} + {K_{1234}} = S({A_4})\\
	S({A_1}{A_2}) = {F_{14}} + {F_{23}} + {F_{24}} + 2{K_{1234}} = S({A_3}{A_4})\\
	S({A_2}{A_3}) = {F_{12}} + {F_{24}} + {F_{34}} + 2{K_{1234}} = S({A_1}{A_4})\\
	S({A_1}{A_3}) = {F_{12}} + {F_{14}} + {F_{23}} + {F_{34}} + 2{K_{1234}} = S({A_2}{A_4})
\end{array}.\end{equation}
We also write down the expression under the previous PEE=CMI bipartite entanglement distillation scheme as a comparison to clarify the similarities and differences. We rename the $F$ under the original bipartite entanglement distllation scheme by $F'$ to distinguish it:
\begin{equation}\label{set2}\begin{array}{l}
	S({A_2}) = F{'_{12}} + F{'_{23}} + F{'_{24}}\\
	S({A_3}) = F{'_{23}} + F{'_{34}} + F{'_{13}}\\
	S({A_1}{A_2}{A_3}) = F{'_{14}} + F{'_{24}} + F{'_{34}} = S({A_4})\\
	S({A_1}{A_2}) = F{'_{14}} + F{'_{23}} + F{'_{24}} + F{'_{13}} = S({A_3}{A_4})\\
	S({A_2}{A_3}) = F{'_{12}} + F{'_{24}} + F{'_{34}} + F{'_{13}} = S({A_1}{A_4})\\
	S({A_1}{A_3}) = S({A_2}{A_4}) \to F{'_{12}} + F{'_{14}} + F{'_{23}} + F{'_{34}} + 2F{'_{13}}
\end{array}.\end{equation}
The last equation is a characterization of the paradox pointed out in section~\ref{sec32}, where we cannot give a reasonable explanation for it in physics. Let us further point out one thing worth noting: from the expressions of (\ref{fij}) and (\ref{sol}), we can see that the number of 2-threads connecting adjacent regions in both schemes is given by mutual information, so they are the same, that is:
\begin{equation}\label{given}
F{'_{12}} = {F_{12}},\;F{'_{14}} = {F_{14}},\;F{'_{23}} = {F_{23}},\;F{'_{34}} = {F_{34}}\;
\end{equation}
Therefore, if we compare (\ref{set1}) and (\ref{set2}), we will immediately find that actually, ${K_{1234}}$ plays the role of ${F{'_ {13}}}$ in (\ref{set1})! More precisely, given (\ref{given}), if we make a replacement:
\begin{equation}\label{repl}
\begin{array}{l}
	F{'_{13}} \to {K_{1234}}\\
	F{'_{24}} \to {F_{24}} + {K_{1234}}
\end{array}.
\end{equation}
then the scheme of (\ref{set2}) will exactly return to the scheme of (\ref{set1})! And the last equation of (\ref{set2}) also becomes physically understandable after the replacement!

\begin{figure}[htbp]     \begin{center}
		\includegraphics[height=8cm,clip]{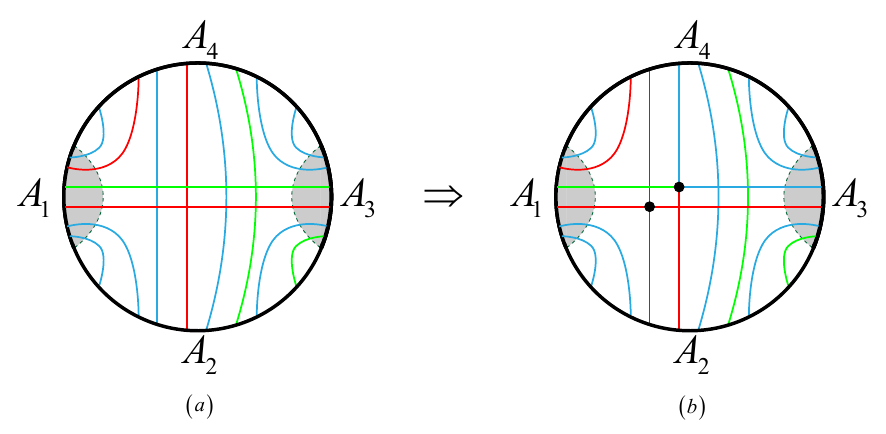}
		\caption{(a)	The original bipartite entanglement distillation scheme of PEE=CMI, where the 2-threads connecting $A_1$ and $A_3$, and 2-threads connecting $A_2$ and $A_4$ are mutually unentangled with each other.
			(b)	The improved distillation scheme after introducing multipartite entanglement, where all ${F'_{13}}$ threads connecting $A_1$ and $A_3$ are actually entangled with an equal number of threads connecting $A_2$ and $A_4$.	}
		\label{fig43}
	\end{center}	
\end{figure}
Let us draw (\ref{repl}) in a graphical way. As shown in figure~\ref{fig43}, the left figure represents the original bipartite entanglement distillation scheme of PEE=CMI, and the right figure represents the improved distillation scheme after introducing multipartite entanglement. There are originally $F{'_{13}}$ (equals 2 in the figure) 2-threads connecting $A_1$ and $A_3$, and ${F'_{24}}$ (equals 4 in the figure) 2-threads connecting $A_2$ and $A_4$. ${F'_{24}}$ is larger than $F{'_{13}}$ numerically because we are considering the case where $A_1$ and $A_3$ are relatively small. The key point is that in this distillation scheme, all these ${F'_{24}}$ and $F{'_{13}}$ threads are mutually unentangled with each other (the overall state is the direct product state of the state of each 2-thread). In the new scheme, all ${F'_{13}}$ threads connecting $A_1$ and $A_3$ are actually entangled with an equal number of threads connecting $A_2$ and $A_4$, so that the overall state of these threads forms the product state of a total of ${K_{1234}}={F'_{13}}$ perfect tensor states. On the other hand, there is still a part of threads connecting $A_2$ and $A_4$, the number of which is ${F'_{24}}-{K_{1234}}$, which exactly provides ${F_{24}}={F'_{24}}-{K_{1234}}$ threads connecting $A_2$ and $A_4$ in the new scheme.

With this understanding, it is now clear how to solve the difficulties in section~\ref{sec321} and the coincidence in section~\ref{sec322}. In section~\ref{sec322}, the reason why we find that in the case where the sizes of $A_1$ and $A_3$ are relatively small, TI $I({A_1}:{A_2}:{A_3})$ is just equal to the number of threads connecting $A_1$ and $A_3$, i.e., ${F'_{13}}$(up to a factor of 1/2 ) (see (\ref{i1})), is because in a more sophisticated distillation scheme, the number of 4-threads that completely replace these 2-threads connecting $A_1$ and $A_3$ is just given by the tripartite information I(${A_1}:{A_2}:{A_3}$) (up to a factor of 1/2) (see (\ref{k1234}). Therefore, in fact, the coincidence in section~\ref{sec322} secretly suggests the rationality of introducing multipartite entanglement.

Now, following the same logic as in section~\ref{sec31}, it is not difficult to write down the description of various physical quantities based on density matrices in this context. Similarly, we can design simplified graphs to describe thread-state models. As shown in figure~\ref{fig41}(b), we define the elementary region as the boundary vertex (marked with a black dot in the figure), and an ``$n$-thread bundle" is simply described by $n$ legs extending from an $n$-valent internal vertex (marked with an empty dot in the figure) connected to the boundary vertex, with each leg assigned the value of the number of $n$-threads contained in the thread bundle to which it belongs. The meaning expressed here is that each $n$-valent component in the simplified graph~\ref{fig41}(b) represents the direct product state of the states of all the $n$-threads contained in the thread bundle. Note that in the simplified graph~\ref{fig41}(b), we present 7 thread bundles, including 6 2-thread bundles ${v_{12}},{v_{13}},{v_{14}},{v_{23}},{v_{24}},{v_{34}}$ with the number of threads they contain being ${F_{12}},{F_{13}},{F_{14}},{F_{23}},{F_{24}},{F_{34}}$, respectively, and a 4-thread bundle ${v_{1234}}$ with the number of 4-threads it contains being ${K_{1234}}$. However, as shown in figure~\ref{fig42}, we find that for different four-partition cases of the quantum system, some thread bundles can disappear, and if we continuously adjust the way of partition, a phase transition will occur. Initially, when the separation distance between ${A_1}$ and ${A_3}$ is relatively large compared to their own sizes, we find that for 2-thread bundles connecting two disconnected regions, ${F_{13}}$ is 0 and ${F_{24}}$ has a nonzero (positive) value. Then, if we adjust it to the case where the sizes of ${A_1}$, ${A_2}$, ${A_3}$, and ${A_4}$ are all identical, we find that ${F_{13}}$ and ${F_{24}}$ are both 0. If the separation distance between ${A_1}$ and ${A_3}$ is relatively small, ${F_{24}}$ becomes 0, and ${F_{13}}$ has a nonzero (positive) value.

Now we write a unified density matrix description of various physical quantities. However, it should be noted that in the three situations mentioned above, the null state corresponding to the disappeared thread bundle needs to be taken in the expression. First, we write down the states associated with these thread bundles by thread/state correspondence:
\begin{equation}
	\left| {{F_{ij}}} \right\rangle  = {(\frac{1}{{\sqrt 3 }}({\left| r \right\rangle _i}{\left| r \right\rangle _j} + {\left| b \right\rangle _i}{\left| b \right\rangle _j} + {\left| g \right\rangle _i}{\left| g \right\rangle _j}))^{ \otimes {F_{ij}}}},
\end{equation}
\begin{equation}
\begin{array}{l}
	\left| {{K_{1234}}} \right\rangle  = (\frac{1}{3}({\left| r \right\rangle _1}{\left| r \right\rangle _2}{\left| r \right\rangle _3}{\left| r \right\rangle _4} + {\left| b \right\rangle _1}{\left| b \right\rangle _2}{\left| b \right\rangle _3}{\left| r \right\rangle _4} + {\left| g \right\rangle _1}{\left| g \right\rangle _2}{\left| g \right\rangle _3}{\left| r \right\rangle _4}\\
	\quad \quad \quad \quad  + {\left| r \right\rangle _1}{\left| b \right\rangle _2}{\left| g \right\rangle _3}{\left| b \right\rangle _4} + {\left| b \right\rangle _1}{\left| g \right\rangle _2}{\left| r \right\rangle _3}{\left| b \right\rangle _4} + {\left| g \right\rangle _1}{\left| r \right\rangle _2}{\left| b \right\rangle _3}{\left| b \right\rangle _4}\\
	\quad \quad \quad \quad  + {\left| r \right\rangle _1}{\left| g \right\rangle _2}{\left| b \right\rangle _3}{\left| g \right\rangle _4} + {\left| b \right\rangle _1}{\left| r \right\rangle _2}{\left| g \right\rangle _3}{\left| g \right\rangle _4} + {\left| g \right\rangle _1}{\left| b \right\rangle _2}{\left| r \right\rangle _3}{\left| g \right\rangle _4}){)^{ \otimes {K_{1234}}_{ij}}}
\end{array}.
\end{equation}
Once again, where, for example, ${\left| r \right\rangle _i}$ represents the state of the ``distilled qutrit'' in elementary region ${A_i}$ indicated by the thread/state rules associated with the state of the red thread, etc. According to the thread/state rules, the distilled state of the entire quantum system, which is the direct product of the states corresponding to all the threads in the locking thread configuration, is now given as:
\begin{equation}
\left| {{\Psi _\Gamma }} \right\rangle  = \left| {{F_{12}}} \right\rangle  \otimes \left| {{F_{13}}} \right\rangle  \otimes \left| {{F_{14}}} \right\rangle  \otimes \left| {{F_{23}}} \right\rangle  \otimes \left| {{F_{24}}} \right\rangle  \otimes \left| {{F_{34}}} \right\rangle  \otimes \left| {{K_{1234}}} \right\rangle .
\end{equation}
Now, we are interested in the expression of PEE under this distillation scheme, based on reduced density matrices. Let us take $	A = {A_1} \cup {A_2} \cup {A_3}$, $\quad \bar A = {A_4}$, then, we can define the PEE by:
\begin{equation}
	{s_A}({A_1}) + {s_A}({A_2}) + {s_A}({A_3}) = S(A),
\end{equation}
or equivalently, using the CMI representation (\ref{equ}):
\begin{equation}
	{I_{14}} + {I_{24}} + {I_{34}} = S(A)
\end{equation}
From (\ref{pee}), we can calculate that under the current distillation scheme:
\begin{equation}
\label{four}\begin{array}{l}
	{I_{14}} \equiv {s_A}({A_1}) = {F_{14}}\\
	{I_{24}} \equiv {s_A}({A_2}) = {F_{24}} + {K_{1234}}\\
	{I_{34}} \equiv {s_A}({A_3}) = {F_{34}}
\end{array}.
\end{equation}
Therefore, following the same logic as in section~\ref{sec3}, we can define the distilled density matrices of PEE, which are respectively:
\begin{equation}
{\rho _{14}} \equiv {\rho _{{A_1} \to A}} = t{r_{{A_4}}}\left| {{F_{14}}} \right\rangle \left\langle {{F_{14}}} \right| = {(\frac{1}{3}{\left| r \right\rangle _1}{\left\langle r \right|_1} + \frac{1}{3}{\left| b \right\rangle _1}{\left\langle b \right|_1} + \frac{1}{3}{\left| g \right\rangle _1}{\left\langle g \right|_1})^{ \otimes {F_{14}}}},
\end{equation}
\begin{equation}
	{\rho _{34}} \equiv {\rho _{{A_3} \to A}} = t{r_{{A_4}}}\left| {{F_{34}}} \right\rangle \left\langle {{F_{34}}} \right| = {(\frac{1}{3}{\left| r \right\rangle _3}{\left\langle r \right|_3} + \frac{1}{3}{\left| b \right\rangle _3}{\left\langle b \right|_3} + \frac{1}{3}{\left| g \right\rangle _3}{\left\langle g \right|_3})^{ \otimes {F_{34}}}},
\end{equation}
and 
\begin{equation}\label{rho24}
{\rho _{24}} \equiv {\rho _{{A_2} \to A}} = t{r_{{A_4}}}(\left| {{F_{24}}} \right\rangle  \otimes \left| {{K_{1234}}} \right\rangle )(\left\langle {{F_{24}}} \right| \otimes \left\langle {{K_{1234}}} \right|).
\end{equation}  	
Thus, we have
\begin{equation}
{I_{ij}} =  - tr({\rho _{ij}}\log_3 {{\rho _{ij}}})
\end{equation}
and correctly obtain (\ref{four}) as a result. Note that from (\ref{rho24}), we can see that ${I_{24}} \equiv {s_A}({A_2})$ will undergo a significant phase transition during the process as shown in figure~\ref{fig42}.

\section{Discussion: Beyond the PEE=CMI scheme}\label{sec5}
\subsection{Limitations of the PEE=CMI scheme}

From the perspective of the thread/state correspondence, the previous sections have made us aware that to fully analyze the entanglement structure of a holographic quantum system, especially in cases involving disconnected regions, multipartite entanglement is actually a necessary and more natural element. On the other hand, the PEE=CMI scheme stubbornly implies a focus on pairs of points. Although we did not conclude that the PEE=CMI scheme is ineffective in the previous section and instead tend to consider the newly introduced multipartite entanglement as a contribution to the CMI, this section will explicitly point out the limitations of the PEE=CMI scheme.

\begin{figure}[htbp]     \begin{center}
		\includegraphics[height=8cm,clip]{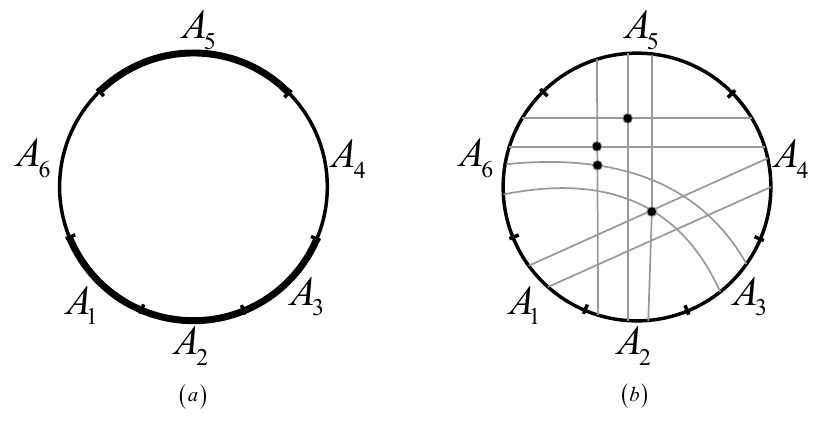}
		\caption{(a) A situation involving six elementary regions, and a disconnected subregion $A = {A_1}{A_2}{A_3} \cup {A_5}$. (b) Various possible non-trivial ways of entangling threads.}
		\label{fig5}
	\end{center}	
\end{figure}
These limitations are essentially related to the characterization of von Neumann entropy for disconnected regions. For example, as an extension of the previous section, we naturally consider a situation involving six elementary regions, as shown in figure~\ref{fig5}(a). Now, let us consider a disconnected subregion $A = {A_1}{A_2}{A_3} \cup {A_5}$, with it complement $\bar A = {A_4} \cup {A_6}$. Then we encounter an obvious difficulty: if we want to consider the contribution of region ${A_2}$ to the von Neumann entropy of $A$, that is, the PEE ${s_A}({A_2})$, we cannot write an appropriate expression for the CMI. The reason is that we cannot uniquely determine which region should be chosen as $L$ in the expression of CMI (\ref{cmi}). When $A_5$ is an empty set, it is obvious that $L$ can be taken as $A_1$, as we commented there, because in this case, $L$ represents the separation distance
 between $A_2$ and $\bar A$, and accordingly, the remaining part of $A$ (after removing $A_2$ and $L$), i.e., $A_3$ can be labeled as $R$. Importantly, due to the properties of pure states, in this case, $L$ can also be taken as $A_3$, because it can equally well characterize the separation distance between $A_2$ and $\bar A$, with $R$ chosen as $A_1$. It is not difficult to prove that the results obtained by these two symmetric choices are exactly the same. However, now that $A_5$ is not an empty set, we face at least two equally reasonable possible choices (which can return to the correct limit when $A_5$ is an empty set), respectively: $L = {A_1}$, $R = {A_3} \cup {A_5}$ or $\quad L = {A_1} \cup {A_5}$, $R = {A_3}$, but it can be verified that these two choices cannot lead to the same expression!

Actually, even if we still consider the four-partition case (see figure~\ref{fig41}), the problem of characterizing the von Neumann entropy of disconnected regions cannot be solved by the PEE=CMI scheme. In the previous section, we found expressions for $I_{14}$, $I_{24}$, and $I_{34}$ by examining the contributions of each component of the region $A = {A_1} \cup {A_2} \cup {A_3}$ to the entropy $S(A)$. If we apply the PEE=CMI scheme to all the connected regions, including ${A_1}$, ${A_2}$, ${A_3}$, ${A_1}{A_2}$, ${A_2}{A_3}$, and ${A_1}{A_2}{A_3}$, we can obtain:
\begin{equation}\label{ifour}
\begin{array}{l}
	{I_{12}} = {F_{12}}\\
	{I_{23}} = {F_{23}}\\
	{I_{34}} = {F_{34}}\\
	{I_{14}} = {F_{14}}\\
	{I_{13}} = {F_{13}} + {K_{1234}}\\
	{I_{24}} = {F_{24}} + {K_{1234}}
\end{array}
\end{equation}
Then, using equation (\ref{equ}), which we will rewrite below for convenience:
\begin{equation}\label{eq}
	{S_{a\left( {a + 1} \right) \ldots b}} = \sum\limits_{i,j} {{I_{ij}}} \;\;\;{\rm{where}}\;i \in \left\{ {a,a + 1, \cdots ,b} \right\},\;j \notin \left\{ {a,a + 1, \cdots ,b} \right\},
\end{equation}
we can correctly determine the von Neumann entropy of all connected regions. However, the problem now is that we cannot use (\ref{eq}) in conjunction with (\ref{ifour}) to correctly calculate the von Neumann entropy of disconnected regions such as ${A_1} \cup {A_3}$ or ${A_2} \cup {A_4}$, since this would result in again:
\begin{equation}\label{fail}
	S({A_1}{A_3}) = S({A_2}{A_4}) = {I_{12}} + {I_{14}} + {I_{23}} + {I_{34}} = {F_{14}} + {F_{24}} + {F_{23}} + {F_{34}},
\end{equation}
which contradicts the last equation in (\ref{set1}). In fact, we have returned to the dilemma in \ref{sec321}.

Our final comment is as follows: the PEE=CMI scheme cannot characterize all the entanglement information of a quantum system, especially the entanglement information of disconnected regions. Let us summarize our efforts and attempts. First, in section~\ref{sec321}, we found that to describe the entanglement structure of a four-partition system in a physically consistent way, it is not possible to simply interpret the CMI as bipartite entanglement, or in other words, it can only be understood as bipartite entanglement in the sense of entanglement distillation. Then, we attempted to solve this problem by introducing four-partite entanglement, and in this case, a self-consistent description of the entanglement structure of the entire four-partition system can be obtained. We then tried to rephrase this description in terms of the CMI language, which led us to (\ref{ifour}) and ultimately to the incorrect result (\ref{fail}).

However, in reality, we don't need to dwell on this. Our efforts do not mean that the concept of PEE is useless. In fact, looking at (\ref{set1}), if we replace the basic objects that characterize entanglement with multipartite entanglement itself, we can still discuss the entanglement structure of the system and indicate how each component in a specified region $A$ (whether it is connected or disconnected) contributes to the von Neumann entropy of $A$ through entanglement with other regions of the system. It just becomes more non-trivial, and we shouldn't expect to be able to simply understand partial entanglement entropy as a simple picture of entanglement between two regions.

\subsection{ Entanglement structure from thread/state correspondence}\label{sec52}

A good way is to further develop the thread/state correspondence picture to characterize the entanglement structure of the holographic quantum system at the level of dividing it into more elementary regions, so as to obtain a more refined understanding of the partial entanglement entropy. We believe that figure~\ref{fig5}(b) provides an inspiring attempt in this direction: we can first construct a locking multiflow for the multi-partite system that satisfies (\ref{equ2}) using the method of section~\ref{sec31}, where each component flow is characterized by a 2-thread bundle. However, we should then find a way, like in figure~\ref{fig43}, to entangle some of the threads belonging to different 2-thread bundles at ``crossing points" to form new $n$-threads, while giving these new $n$-threads a more refined distilled state, such as the perfect tensor state in section~\ref{sec4}.\footnote{Here, it is appropriate to add a comment. Actually, the scheme of constructing the complete entropy vector using ``K-basis" described by equation (\ref{set}) is universal for any number of elementary regions (monochromatic regions)~\cite{He:2019ttu}. However, as pointed out in~\cite{Harper:2022sky}, interpreting a general ${K_{12 \cdots (2s)}}$ component as the number of threads in a 2s-hyperthread bundle faces a problem: there is a possibility of negative values, which forces us to imagine $n$-threads with a ``negative" number of threads.}

Figure~\ref{fig5} is a simplified schematic diagram of this idea, showing various possible non-trivial ways of entangling threads: for example, it can be imagined that three threads belonging to different three thread bundles intersect at a point in the bulk, and then these three threads are integrated and entangled to form a 6-thread. More generally, three threads can be allowed to intersect at multiple points rather than just one, forming a 6-thread with internal edges.

In summary, we have clearly stated a non-trivial problem: how to construct these thread entanglement schemes more generally when dividing the quantum system into more and more elementary regions, and how to properly characterize the distilled state dual to these $n$-thread, so that not only the correct von Neumann entropy of these elementary regions themselves, but also that of every connected and disconnected composite region involved can be given by the distilled density matrices in this framework? We believe that stating this problem clearly is a meaningful step in exploring the entanglement structure of holographic quantum systems, and our approach provides an inspiring direction. However, we will leave the more specific schemes to future work.

\section{Conclusions and discussions}\label{sec6}
This article is a reflection and commentary on the idea of partial entanglement entropy (PEE) for fine-graining the holographic entanglement entropy. It is apparent that the idea of PEE is a typical product of the philosophy of ``the whole equals the sum of its parts", while quantum entanglement inherently possesses the peculiar feature of ``the sum of its parts not equaling the whole". Specifically, we focus on the CMI (conditional mutual information) formulation of PEE. As shown in the kinematic space representation~\cite{Czech:2015kbp,Czech:2015qta,Rolph:2021nan,Lin:2022agc,Lin:2022flo}, a more natural way to express a CMI of the holographic quantum system is to relate it to a pair of elementary regions in the original space representation, which is dual to a wedge-shaped region in the kinematic space. Therefore, our topic can also be reduced to how to understand the actual physical meaning of such kind of correlation of ``region pair". Meanwhile, this work also provides a tentative definition of PEE based on density matrices. To our knowledge, the density matrix-based definition of PEE has not yet been established.

Our work is inspired by the concept of locking bit thread configurations~\cite{Freedman:2016zud,Cui:2018dyq,Headrick:2017ucz,headrick2022crossing} and the subsequently developed thread-state correspondence~\cite{Lin:2022agc,Lin:2022flo,Harper:2022sky}. Since the CMIs one-to-one match the component flow fluxes in a locking bit thread configuration~\cite{Lin:2021hqs}, our first direct result is to interpret CMI as bipartite entanglement obtained from the distillation of the holographic quantum system. However, we discovered a paradox related to disconnected regions. Through the picture of thread-state correspondence, we found that even in the sense of entanglement distillation, if CMI is simply understood as bipartite entanglement, it is impossible to consistently characterize the entanglement entropies of a series of connected and disconnected regions at the same time. This led us to overcome this difficulty by introducing multipartite entanglement. In the picture of thread-state correspondence, this amounts to the introduction of $n$-hyperthreads~\cite{Harper:2021uuq,Harper:2022sky}. Each $n$-hyperthread corresponds to a perfect tensor state. Through this introduction of $n$-thread/perfect tensor state correspondence, we realized that in order to characterize the entanglement information of disconnected regions, the scheme containing merely ``bipartite correlation"(i.e., the so-called ``component flows'') must be further refined, such as by modifying the states of different bundles of 2-threads (which characterize the correlation between two elementary regions) from direct product states to entangled states.

The introduction of multipartite entanglement can indeed solve the problem of characterizing the entanglement entropies of disconnected regions at least in the sense of entanglement distillation. However, we still found in the end that trying to integrate this multipartite entanglement into the PEE=CMI scheme would still lead to a contradiction at the whole level, as shown in equation~(\ref{fail}). Therefore, we finally re-examined the limitations of the idea of ``the whole equals the sum of its parts". The idea of decomposing the entanglement entropy of a region $A$ into the contributions of its various components ${A_i}$ may not be globally applicable to analyzing the entanglement structure of the entire system and should be viewed as an effective description of the entanglement structure of local subregions. On the other hand, the concept of CMI, which represents the ``region pair correlation", is still useful in the holographic duality, and our exploration suggests that it can be used as the first step of the holographic entanglement distillation scheme that can characterize a set of entanglement entropies for all elementary regions and all connected composite regions. However, we then need to entangle the different bundles of ``pair correlations" to form more refined multipartite entanglement in order to achieve the ultimate goal of characterizing all disconnected regions. In the process of achieving this goal, the idea of the thread-state correspondence is very useful and enlightening. In addition to developing the idea of $n$-thread/quantum error-correcting code state duality, in fact, in our next series of preparatory work, we will try to develop the thread-state duality along another path, that is, to introduce the concept of quantum superposition of locking multiflow configuration states, in order to further solve the problems indicated in this paper.

\section*{Acknowledgement}
We would like to thank Ling-Yan Hung and Yuan Sun for useful discussions.

\bibliographystyle{ieeetr}
\bibliography{distilledref}

\end{document}